# Electronic structure of superconducting Lu$_2$Ni$_3$Si$_5$ and its reference compound Y$_2$Ni$_3$Si$_5$ by *ab initio* calculations

M. Samsel-Czekała* and M.J. Winiarski

*Institute of Low Temperature and Structure Research, Polish Academy of Sciences, P.O. Box 1410, 50-950 Wrocław 2, Poland*

Electronic structures of orthorhombic ternary nickel silicides: superconducting Lu$_2$Ni$_3$Si$_5$ and its non-superconducting counterpart, Y$_2$Ni$_3$Si$_5$, have been calculated employing the fully-relativistic and full-potential local-orbital method within the density functional theory. Our investigations were focused particularly on the band structures and Fermi surfaces (FSs), being very similar for both ternaries. It appears that their FSs exist in four bands and contain electronlike and holelike three-dimensional sheets and small pockets, which suggests a presence of two- or even multi-band superconductivity (SC) in Lu$_2$Ni$_3$Si$_5$. The main difference between both systems is that only in Lu$_2$Ni$_3$Si$_5$ small electron FS pockets occur around the Γ point. It allows for arising BCS-like SC in this compound, as was deduced from previous heat-capacity measurements, while no sign of SC has been detected (at least down to 2 K) in Y$_2$Ni$_3$Si$_5$. In the latter system, a possible formation of a ferromagnetic ground state, which usually tends to destroy SC, has been excluded by our calculations.

## 1. Introduction

Wide interest in nickel-based (oxy)pnictide superconductors mainly of 1111, 122 and 344-type families, known in the literature, e.g. [1,2,3], containing the similar Ni-*X* atomic layers, where *X* is a pnictide, chalcogenide or boron atom, motivated us to investigate also other superconducting Ni-based materials possessing such layers with *X* = Si. Nickel-pnictide-like structures are usually strongly anisotropic, forming quasi-two-dimensional (Q2D) systems. They are built from positively charged layers of atoms of alkaline or rare-earth metals and negatively charged layers containing nickel and non-metallic atoms. Although in these families the superconducting transition temperatures, $T_C$, are rather low, usually below 5 K, they are widely investigated nowadays in analogy to the high-$T_C$ iron-based rare-earth pnictides, like SmFeAsO$_{1-x}$F$_x$ and Sr$_{1-x}$Sm$_x$FeAsF, reaching the highest $T_C$ of 55 K due to doping [4]. Interestingly, among Ni-based superconductors, contrary to Fe-based ones, rather three-dimensional (3D) rare-earth borocarbides, achieve $T_C$ maximum of 16.5 K [5]. So far, a mechanism of high-temperature superconductivity (SC) has been not recognized. Therefore, a comparison of electronic structures of similar systems, based on nickel or iron atoms, might be crucial in understanding an SC phenomenon in such a class of materials.

The considered in this work family of ternary rare-earth nickel silicides adopts an orthorhombic structure of the U$_2$Co$_3$Si$_5$ – type (*Ibam*, space group no. 72) containing the Ni-Si layers lying much closer to one another, along the *c* axis, than those in other nickel-silicides, as e.g. La$_3$Ni$_4$Si$_4$ [6,7]. It is interesting that most of the $R_2$Ni$_3$Si$_5$-family members exhibit antiferromagnetic phase transitions at temperatures $T_N \leq 30$ K and the magnetic moments, originating only from the rare-earth atoms because the nickel atoms remain non-magnetically ordered (NMO) [8-14]. In antiferromagnetically ordered (Pr;Dy;Ho)$_2$Ni$_3$Si$_5$ systems, an existence of also short-range ferromagnetic (FM) correlations has been postulated, based on observations of an anomalous behavior of magnetoresistivity [15]. Up to now, among members of the $R_2$Ni$_3$Si$_5$ family, the superconductivity phenomenon has been detected only in Lu$_2$Ni$_3$Si$_5$ with $T_C$ = 2.1 K, being metallic but NMO at least above the $T_C$ [16]. Based on heat-capacity measurements, the SC mechanism in this material was interpreted as being rather weak-coupling BCS-like ($\Delta C/\gamma T$ = 1.1 at T = $T_C$) [16]. Similarly to Lu$_2$Ni$_3$Si$_5$, the isostructural and isoelectronic Y$_2$Ni$_3$Si$_5$ system was found to be NMO (down to 4.2 K) and its electronic transport properties (measured down to 2 K), though showed also a metallic state, did not reveal any evidence of SC [17]. If the ground state of Y$_2$Ni$_3$Si$_5$ becomes, in fact, of the NMO kind, this compound may be considered as a candidate to be superconducting like Lu$_2$Ni$_3$Si$_5$. This expectation is inferred from an analogy to the family of tetragonal ternary iron silicides, in which only (Lu;Y;Sc)$_2$Fe$_3$Si$_5$, being just NMO members, exhibit SC [18]. Based on the analysis of heat capacity measurements of Lu$_2$Fe$_3$Si$_5$, two-band SC having a weak-coupling BCS-like character was suggested [19].

Up to our best knowledge, no electronic-structure calculations results for both considered here ternaries, Lu$_2$Ni$_3$Si$_5$ and Y$_2$Ni$_3$Si$_5$, have been reported in the literature so far. Thus, in this paper, such results are presented for both compounds, with the main focus on their Fermi surfaces (FSs) analyzes at ambient and higher pressure from the point of view of possible SC mechanisms. Since as yet no experimental data have been available below 2.0 K, a possibility of a magnetically ordered ground state in Y$_2$Ni$_3$Si$_5$ was examined by us as well.

## 2. Computational methods

Electronic structure calculations of (Lu,Y)$_2$Ni$_3$Si$_5$ have been performed with a modern full-potential local-orbital (FPLO) method [20]. The Perdew-Wang form of the local (spin)-density approximation [L(S)DA] of exchange-correlation functional [21] was employed in both fully and scalar relativistic modes, the latter without an inclusion of the spin-orbit (SO) coupling. Experimental x-ray diffraction values of lattice parameters of the unit cell (u.c.) possessing the *Ibam* symmetry are as follows: $a = 0.9604$, $b = 1.1014$, $c = 0.5512$ nm for Lu$_2$Ni$_3$Si$_5$ [16] and $a = 0.95651$, $b = 1.11284$, $c = 0.56453$ nm for Y$_2$Ni$_3$Si$_5$ [22]. It should be mentioned that according to Ref. [16] the crystal structure of the Lu-based compound, probed on the powder sample, could not be assigned unambiguously and some slight distortion to the monoclinic variation (*C2/c*) of the orthorhombic (*Ibam*) structure could not be excluded. They were used as initial parameters in further optimization of the u.c. volumes at ambient and higher pressure by minimizing total energy but keeping the internal atomic coordinates (given below) fixed, which yielded some approximation of a pressure dependence of crystal and electronic structures. Here u.c. is equivalent to double formula units (f.u.). The same experimental atomic positions in u.c. as those obtained by the single-crystal x-ray data refinement for Y$_2$Ni$_3$Si$_5$ [22] were taken for both studied systems. This assumption is justified taking into account the fact that the isoelectronic Y and Lu atoms occupy equivalent positions in the crystal structure and the experimental atomic positions of Y$_2$Ni$_3$Si$_5$ and e.g. those of Sm$_2$Ni$_3$Si$_5$ [14] differ only negligibly, despite the considerable (compared to the pair of Y and Lu atoms) disparity in size between the Y and Sm atoms. The crystal structure is visualized in Fig. 1 where the used experimental atomic positions of Y$_2$Ni$_3$Si$_5$ [22] are as follows: Y (Lu) in (8j): (0.2632, 0.3691, 0); Ni(1) in (4b): (1/2, 0, 1/4); Ni(2) in (8j): (0.1123, 0.1342, 0); Si(1) in (8g): (0, 0.2663, 1/4); Si(2) in (8j): (0.3475, 0.1071, 0); Si(3) in (4a): (0, 0, 1/4). The following valence-basis sets were selected in our calculations: the Y: 4s4p;5s5p4d; the Lu: 5s5p4f;6s6p5d; the Ni: 3s3p;4s4p3d, and the Si: 2s2p;3s3p3d states. Total energy values of considered systems were converged with accuracy to ~1 meV for the 16x16x16 (621 points) *k*-point mesh in the non-equivalent part of the Brilouin zone (BZ). A possible FM ground state in Y$_2$Ni$_3$Si$_5$ was examined by standard spin-polarized, scalar relativistic self-consistent LSDA calculations. For this purpose, the fixed spin moment (FSM) method [23], implemented in the FPLO code, was utilized as well.

## 3. Results and discussion

For both (Lu;Y)$_2$Ni$_3$Si$_5$ compounds, the theoretically optimized volumes of u.c., V$_{calc}$, amount to about 95.9% of their experimental volumes, V$_{exp}$. Results of electronic structure calculations, presented here, were obtained for the above V$_{calc}$ and they do not differ considerably from those determined for V$_{exp}$. The total and partial DOSs of both studied systems, based on the fully relativistic approach, are plotted in Figs. 2 and 3. As seen in these figures, their overall shapes are similar for both compounds, differing mainly in the additional presence of two narrow peaks of the Lu 4f electron states. Our fully relativistic calculations for Lu$_2$Ni$_3$Si$_5$ indicate that these are the SO split Lu 4f$_{5/2}$ and 4f$_{7/2}$ peaks, located at 5.4 and 3.9 eV below the Fermi level (E$_F$), respectively (see upper inset to Fig. 2). As expected, based on the scalar relativistic calculations, one was able to get only one high peak of the 4f electrons occurring at about -4.5 eV (displayed in upper inset to Fig. 2). It is interesting to note that by employing *ab initio* calculations also in the scalar relativistic approach but for some other ternary lutetium and nickel compounds, e.g. LuNiAl, a similar single peak of the Lu 4f electrons around -5 eV was obtained [24]. It is worth underlining that the Lu 4f electrons in $Lu_2Ni_3Si_5$ are localized well below $E_F$ and they do not contribute to the DOS around the Fermi level.

In both investigated compounds, DOSs at the Fermi level exhibit some pronounced deeps and, hence, they have at $E_F$ relatively low values [~2.5 electrons/(eV*f.u.)], being typical of rather weak metals and numerous low-carrier superconductors (see e.g. [25]). Furthermore, the densities around $E_F$ are dominated by almost equal electron contributions coming from all three kinds of constituent atoms, forming metallic bonds, *i.e.* the Lu 5d (or Y 4d), Ni 3d and Si 3p electrons. These electron orbitals can be just responsible for the occurrence of SC. As seen in Fig. 3, there is substantial Lu/Y d character at $E_F$. As a consequence, the electron-phonon interaction, which may arise SC, could be different between these two materials. It can be explained by the fact that the Y atom is much lighter than the Lu atom, thus all things being equal its phonons would be at different frequencies. The DOSs below $E_F$ contain mainly the Ni 3d electrons creating a broad peak centered at about -3.0 eV and ranging from about -6 to 3 eV. Moreover, there are about five times smaller-intensity contributions originating from the Si 3p electrons, with their maxima at about -4 eV and spreading from about -12 eV to 5 eV. In turn, contributions from the Lu 5d (or Y 4d) electrons, occurring in the same energy region, have about three times lower values at their maxima being situated around -2 eV. In all cases, calculated contributions to DOS originating from two non-equivalent (in u.c.) positions of nickel atoms turned out to be almost equal and, similarly, contributions coming from all three different atomic sites of silicon are comparable to one another. Only slight differences have been observed in electron population analysis given below.

Calculated electronic occupation numbers, $N_{calc}$, in the $(Lu,Y)_2Ni_3Si_5$ compounds compared to those for the free atoms, $N_{at}$, are collected in Table I. As this table indicates, the electron populations in the Ni and Si atoms are the same (within accuracy to 0.1) in both systems. In turn, there are small differences ($\leq 0.2$) of these populations within each silicide, depending on atomic positions in u.c., from which given electrons originate. The Lu 4f states, having $N_{calc} \sim 14.0$, can be regarded as rather well localized and weakly hybridized with other states. Which is important, the numbers of all kinds of hybridized d electrons, creating a metallic bond, are substantially increased. For the Lu 5d and Y 4d states, $N_{calc} \cong 2N_{at}$ and for the Ni 3d electrons, $N_{calc} \cong N_{at} + 1$ and, finally, the Si 3d states (not present in free atoms) do occur in the studied ternaries. Moreover, in all constituent atoms their s electrons have $N_{calc} < N_{at}$ while valence p electrons in Lu or Y and Ni (non-existing in free atoms) occur with enhanced values, $N_{calc} \cong 0.8$.

The deepest valence bands, dominated by the Si 3s electrons, range down to about 12.5 eV below $E_F$ for both compounds. Except for the discussed presence of the Lu 4f electrons, a pronounced difference has been observed in the band region from -6.8 to -6.6 eV where only the yttrium-based compound has a distinct and narrow energy gap, visible in Fig. 3. Otherwise, in the lutetium-based system, due to an influence of the Lu 4f on Si 3p electrons, there is only a pseudo-gap at about -7.0 eV as seen in Fig. 3.

In Figs. 4 and 5 band energies, $\varepsilon_n(\mathbf{k})$, are plotted for the system containing Lu and Y atoms, respectively, in the vicinity of $E_F$. As seen in these figures, band structures of isoelectron Lu- and Y-based systems are very similar to each other near $E_F$, differing mainly along the ΓY line. Among them, only superconducting $Lu_2Ni_3Si_5$ has electron pockets along this line, one is centered at the Γ point and the other is distant by about 2/3|ΓY| from the Γ point. These pockets coming from the 329th band have ellipsoidal shape, which is clearly seen in Fig. 6, where FS sheets of both compounds are visualized. Taking into account that $Lu_2Ni_3Si_5$ was inferred from the experimental heat-capacity data of Ref. [16] to be a weak-coupling BCS-like superconductor, these ellipsoidal electron pockets can be mainly responsible for SC. Close similarity of electronic structures in both investigated compounds causes some paradox. If these subtle FS pockets did not play a crucial role in creating superconductivity in $Lu_2Ni_3Si_5$, it would be unclear why as yet this phenomenon has been not observed experimentally (at least down to 2.0 K) in $Y_2Ni_3Si_5$.

Pressure dependence of band structure near Fermi energy for $Y_2Ni_3Si_5$ and $Lu_2Ni_3Si_5$ has been investigated by us focusing mainly on the electronlike pockets lying along the ΓY line. As one can deduce from Figs. 4 and 5, performed simulations of increasing pressure by taking smaller u.c. volumes, lead to growing the pseudo-gap size near the Fermi level and also to a suppression of the pockets. Furthermore, if the pockets, indeed, are substantial in superconductivity, any external pressure would suppress SC in $Lu_2Ni_3Si_5$ as well. Oppositely, an effect of internal "negative pressure", due to e.g. chemical doping, would be able to enlarge or create the pockets. As a consequence, SC might be enhanced or arisen in $Lu_2Ni_3Si_5$ or $Y_2Ni_3Si_5$, respectively.

Fermi surfaces of $Lu_2Ni_3Si_5$ and $Y_2Ni_3Si_5$, as demonstrated in Fig. 6, exist in four Kramers double-degenerated conduction bands, numbered as (325-331) and (269-275), respectively, and exhibit a three-dimensional (3D) character. For both compounds, FS sheets coming from bands 327 for Lu- and 271 for Y-based compounds are large and complex possessing a holelike character. There are also big electronlike sheets in bands 329 for Lu- and 273 for Y-based systems having some nesting properties along the ΓZ direction. In addition, small pockets exist in the 325 (or 269) bands, having a holelike character as well as electronlike pockets occur in the 331 (or 275) bands in the Lu- (or Y-) based systems. However, the discussed above electronlike pockets in the 329th band of $Lu_2Ni_3Si_5$ are bigger and more stable than the former and, hence, they seem to play the most important role in a formation of the superconducting state.

The presented here FSs in $(Lu;Y)_2Ni_3Si_5$ have the 3D multi-band characters possessing large and complex FS sheets as well as small pockets, being similar to those observed in the Ni-based multi-band strong-coupling BCS-like $(Lu;Y)Ni_2B_2C$ superconductors [26,27]. Such FSs differ considerably from those observed in the Ni-based 1111-type family, where they possess quasi-2D cylindrical FS sheets [1] like those found in the famous Fe-based 1111 or 122-type superconductors [25,28].

Finally, the ground state of $Y_2Ni_3Si_5$, determined in our standard LSDA calculations, is NMO. This feature is also confirmed by the FSM calculations, revealing the minimum in the total energy, as shown in Fig. 7, for zero ordered magnetic moment per u.c. (and also per each atomic position). Therefore, $Y_2Ni_3Si_5$ is predicted to be paramagnetic down to the lowest measured temperatures, as is also the case of $Lu_2Ni_3Si_5$.

## 4. Conclusions

The band structures of superconducting $Lu_2Ni_3Si_5$ and its non-superconducting reference $Y_2Ni_3Si_5$ system have been studied from the first principles by modern FPLO code. For both compounds, except for subtle electronlike pockets, located along the ΓY line and occurring in the Fermi surface of solely $Lu_2Ni_3Si_5$, as well as the presence at $E_F$ of d states originating from different Lu or Y atoms, which might lead to different electron-phonon interactions responsible for SC, the other features of their electronic structure, particularly near the Fermi level, are almost identical. Hence, one can expect that the FS pockets are crucial in arising superconductivity, as yet detected only in the case of $Lu_2Ni_3Si_5$. The FSs of both systems originate from four bands and contain electronlike and holelike three-dimensional sheets as well as different kinds of small pockets. Thus, this feature may favor in $Lu_2Ni_3Si_5$ two- or even multi-band BCS-like superconductivity. Our calculations also indicate that an external pressure applied to $Y_2Ni_3Si_5$ would not make any further improvement in a resemblance between its FS and that in $Lu_2Ni_3Si_5$. Furthermore, the computed ground state of $Y_2Ni_3Si_5$ is non-magnetically ordered, contrary to magnetic states observed in other non-superconducting compounds among the $R_2Ni_3Si_5$ family.

**Acknowledgments**

The National Center for Science in Poland is acknowledged for financial support of Project No. N N202 239540. The Computing Center at the Institute of Low Temperature and Structure Research PAS in Wrocław is acknowledged for the use of the supercomputers and technical support.

Canfield, B.N. Harmon, and A. Kaminski, Phys. Rev. Lett. **101**, 177005 (2008).

**Table 1.** Calculated occupation numbers (per given orbital at 1 atomic position) in the $(Lu,Y)_2Ni_3Si_5$ systems, $N_{calc}$, compared to those in free atoms, $N_{at}$, with accuracy to ± 0.1. Note that $N_{calc}$ for the same electronic states in the Ni and Si atoms are the same in both ternaries but it is varying slightly (± 0.2) depending on atomic positions in given u.c., from which the electrons originate.

| state | $N_{at}$ | $N_{calc}$ | state | $N_{at}$ | $N_{calc}$ |
|---|---|---|---|---|---|
| Lu 4f | 14 | 14.0 | Ni 3d | 8 | 8.8-8.9 |
| Lu 5d | 1 | 2.1 | Ni 4s | 2 | 0.6 |
| Lu 6s | 2 | 0.6 | Ni 4p | 0 | 0.7-0.9 |
| Lu 6p | 0 | 0.8 | Si 3s | 2 | 1.2-1.3 |
| Y 4d | 1 | 2.2 | Si 3p | 2 | 2.0-2.2 |
| Y 5s | 2 | 0.5 | Si 3d | 0 | 0.2-0.4 |
| Y 5p | 0 | 0.7 | | | |

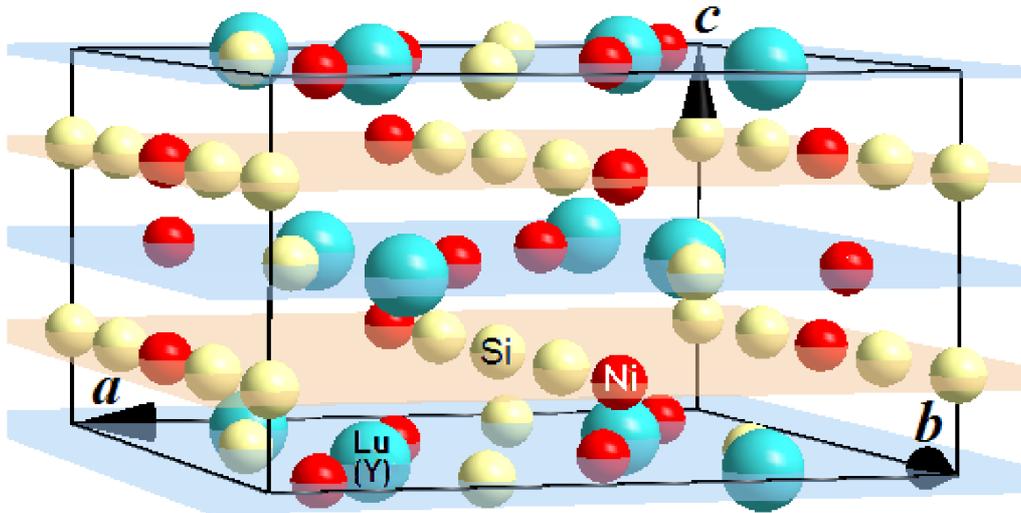

**Fig. 1** Orthorhombic *Ibam* crystal unit cell of $(Y;Lu)_2Ni_3Si_5$ compounds of the $U_2Co_3Si_5$-type (no. 72).

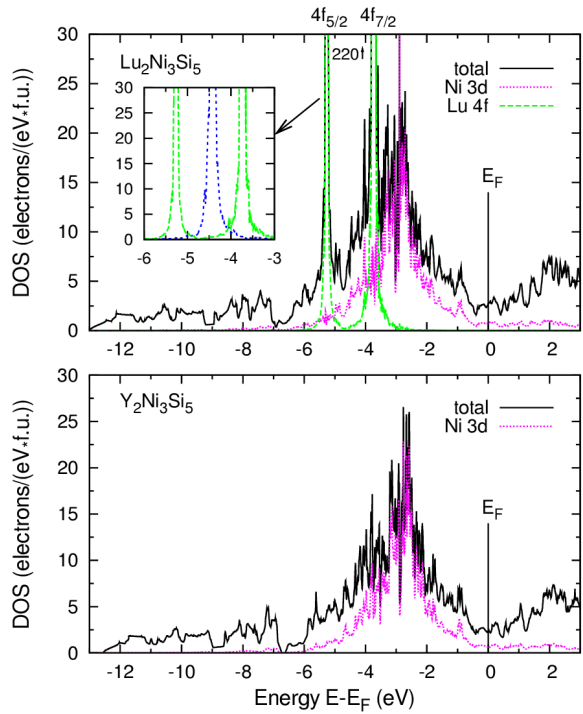

**Fig. 2** Total and partial (per Ni 3d and Lu 4f electron orbitals) DOSs in $R_2Ni_3Si_5$ for $R$ = Lu and Y, obtained in the fully relativistic mode. Upper inset shows the Lu $4f_{5/2}$ and $4f_{7/2}$ peaks on expanded scale. In the middle of this inset, there is hypothetical single peak of the Lu 4f electrons, yielded in the scalar relativistic mode (without SO splitting).

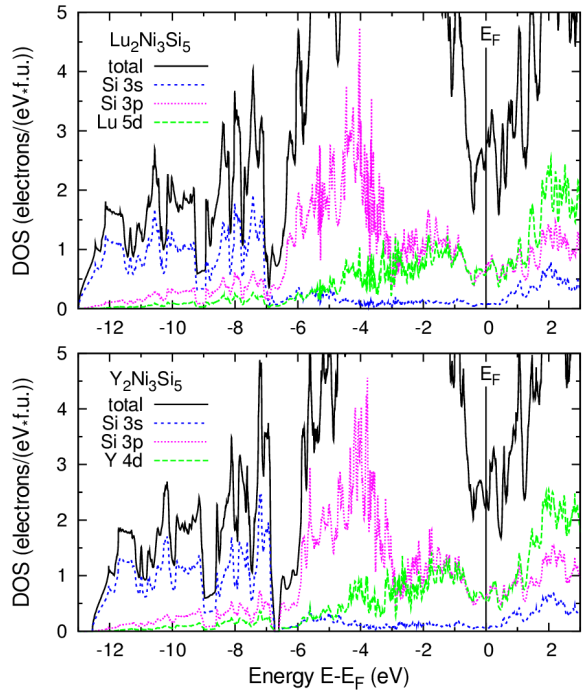

**Fig. 3** Same as in Fig. 2 but for different partial DOSs displayed on expanded DOS scale.

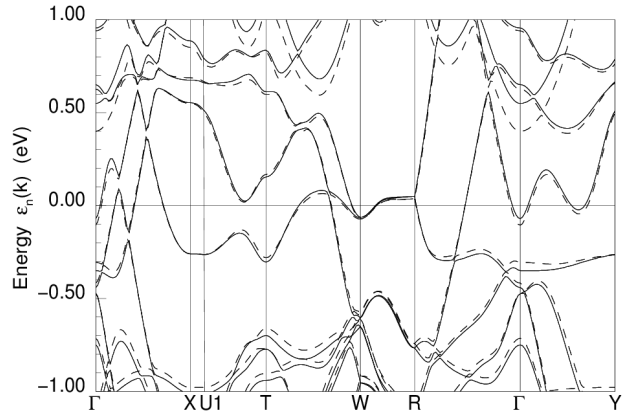

**Fig. 4** Fully relativistic band energies, $\varepsilon_n(\mathbf{k})$, in $Lu_2Ni_3Si_5$ calculated for the LDA equilibrium, $V_{calc}$(= 95.9% $V_{exp}$), (solid line) and experimental, $V_{exp}$, (dashed line) u.c. volumes.

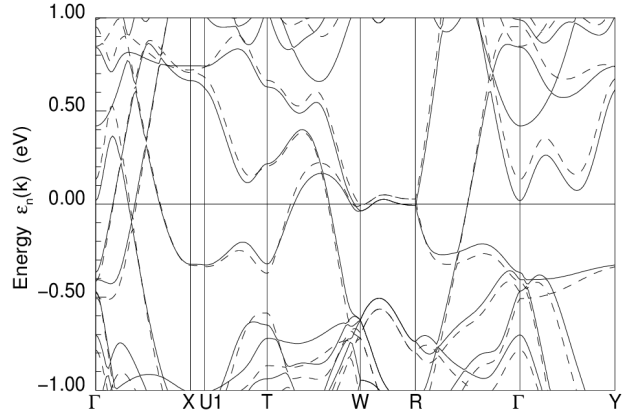

**Fig. 5** Fully relativistic $\varepsilon_n(\mathbf{k})$ in $Y_2Ni_3Si_5$, calculated for the LDA equilibrium, $V_{calc}$ = 95.9% $V_{exp}$, (solid line) and for u.c. volume under pressure, $V_p$ = 85.7% $V_{exp}$, (dashed line).

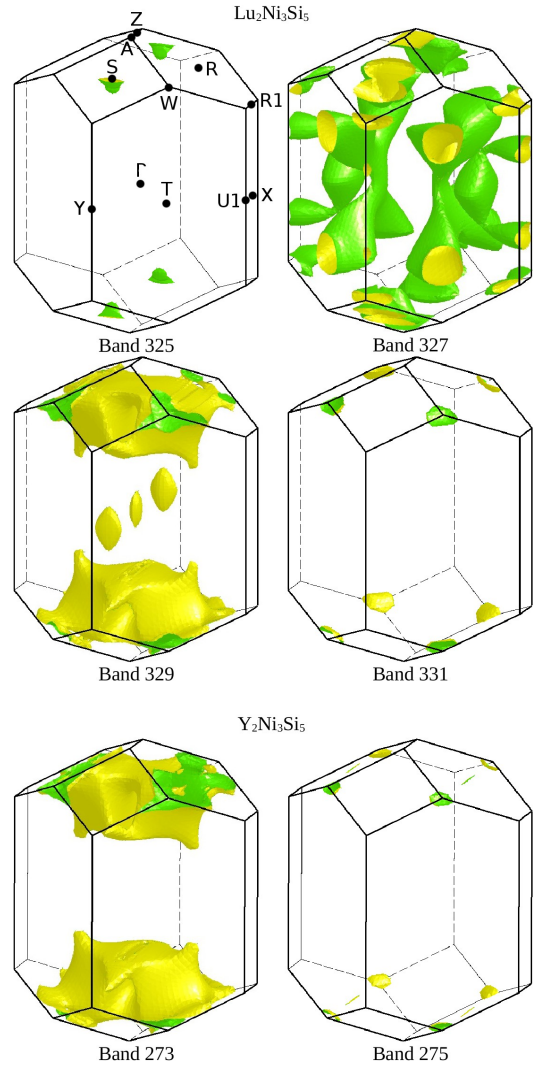

**Fig. 6** Fully relativistic FS sheets calculated (LDA) for $Lu_2Ni_3Si_5$, existing in four different Kramers double-degenerated bands (numbered as 325-331). Since computed FS sheets of $Y_2Ni_3Si_5$ originating from two lower bands (269 and 271) are almost identical to those coming from the bands 325 and 327 in $Lu_2Ni_3Si_5$, only FS sheets from two upper bands (273 and 275) are displayed here.

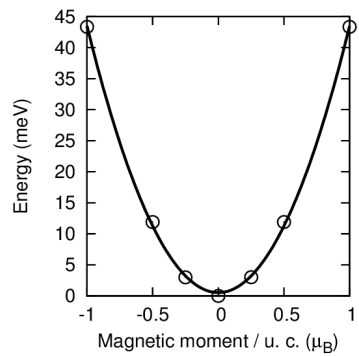

**Fig. 7** Dependence of the total energy on the spin moment per u.c. obtained in the FSM calculations for $Y_2Ni_3Si_5$.